\documentstyle[epsfig]{aipproc}

\begin{document}

\def\d{{\rm d}}
\def\half#1{{1 \over 2} }
\def\gf{\gamma_5}
\def\Tr#1{\rm ~Tr}
\def\inclus#1#2#3 {#1 {\d^3 #2 \over \d^3 #3}}
\def\incltwo#1#2#3#4#5 {#1\, #2 {\d^6 #3 \over \d^3 #4\, \d^3 #5 }}
\def\ga{\gamma_{\alpha}}
\def\gb{\gamma_{\beta}}
\def\gt{\gamma_{\tau}}
\def\gs{\gamma_{\sigma}}
\def\rin{\rho_{\rm in}}
\def\rout{\rho_{\rm out}}
\def\posup{e^+_{\uparrow}}
\def\qup{q_{\uparrow}}
\def\dLL{$\hat d_{LL}$ }

\title{Polarization transfer in SIDIS \\ for $\Lambda$ production$^{\dag}$
}
\author{ Yuri Arestov}
\address{Institute for High Energy Physics\\
142284 Protvino, Moscow Region, Russia}
\maketitle
\begin{abstract}
Polarization transfer from the longitudinally polarized positron
to the final-state quark  is considered for the underlying  QCD
subprocesses giving rise to the reaction 
$e^+_{\uparrow} + p \to  e^+ + \Lambda_{\uparrow} + X $ 
at the HERMES energy.
\end{abstract}
%
Measuring the longitudinal $\Lambda$ polarization in the semi-inclusive
deeply inelastic scattering (SIDIS) 
\begin{eqnarray}
e^+_{\uparrow} + p \to  e^+ + \Lambda_{\uparrow} + X \, ,
\label{sidis}
\end{eqnarray}
that is under study in the experiment HERMES at DESY,
may give important information for the determination of the
twist-2 polarized fragmentation function $G_1 (z,\mu )$.
The latter can be found through the relation
\begin{eqnarray}
   D_{LL} \sim F_q (x)\cdot\hat\sigma
\cdot\hat d_{LL} \cdot G_1 (z,\mu )\, , \label{G1}
\end{eqnarray}
where $F_q(x)$ is the unpolarized parton (quark/gluon) distribution
function in the proton
and the polarization transfer $D_{LL} $ is defined as
\begin{equation}
   D_{LL}=\frac{\sigma_{++}+\sigma_{--}-\sigma_{+-}-\sigma_{-+}} 
{\sigma_{++}+\sigma_{--}+\sigma_{+-}+\sigma_{-+}} \,\equiv \, 
\frac{{\rm d}\Delta\sigma}{{\rm d}\sigma} \, ,      \label{Dll}
\end{equation}
and it can be measured in the experiment. The hatted entities in (\ref{G1})
relate to the underlying QCD subprocess. The subscripts in $\sigma$'s
in (\ref{Dll}) denote the positron and $\Lambda$ helicities.

  The published HERMES measurement in the reaction (\ref{sidis}) 
is  $D_{LL}=0.11\pm 0.17\pm 0.03 $\cite{exper}.

As is seen from (\ref{G1}), it is very instructive to study
\dLL for the purpose of the data handling and analysis. The 
parameter can also be expressed as \dLL\,=\,
${P^z_{out}/ P^z_{in}}$ 
------------------------------------------------------- \\
{\bf $\dag$ Talk given at IX Workshop on High Energy Spin Physics, 
 Dubna, Aug. 2001} \\
where $P$'s are the longitudinal 
components of the initial and final polarization vectors.

\begin{figure}[h]
\begin{minipage}{5.5cm}
\fbox{\epsfig{file=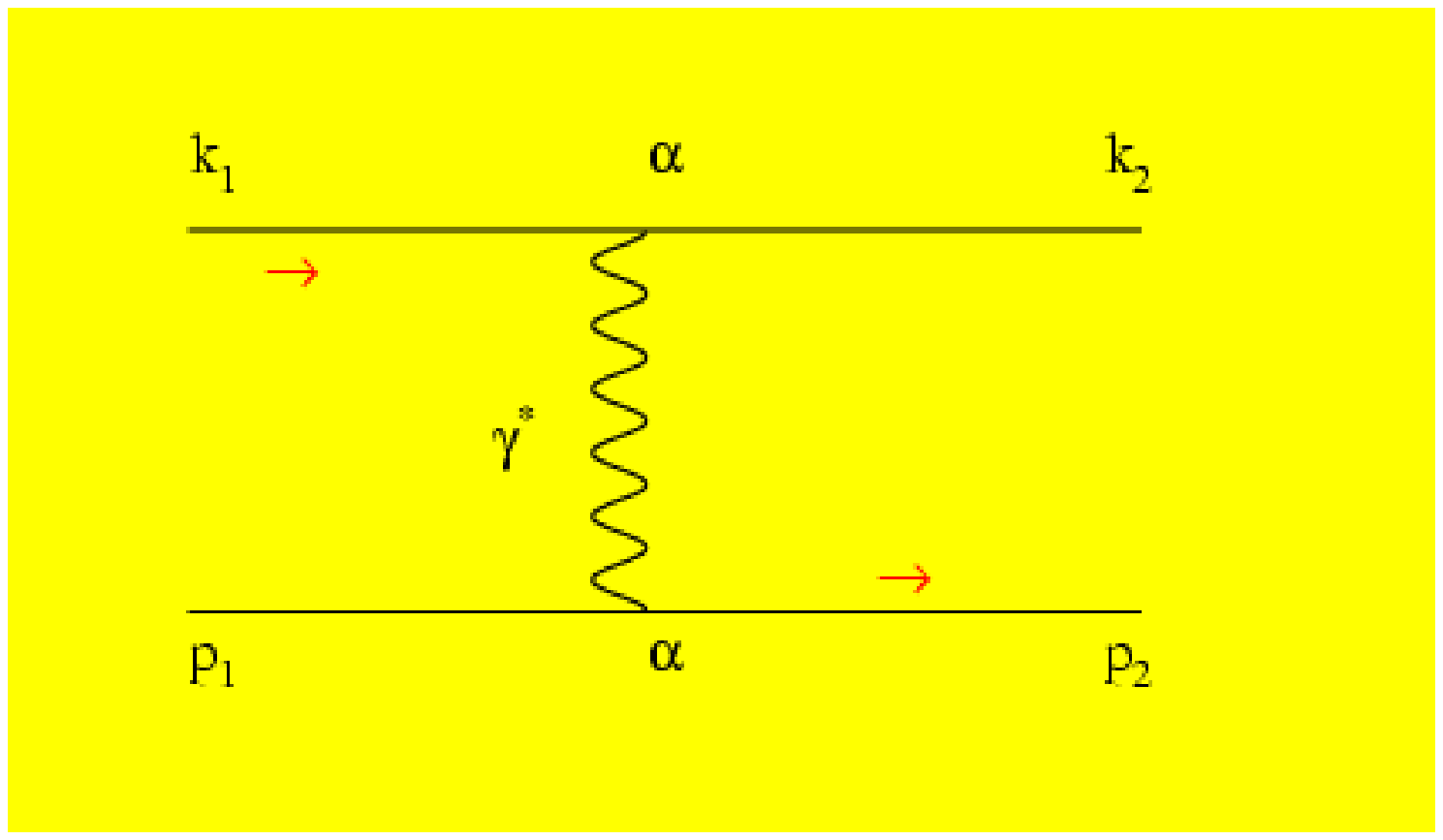,width=5.5cm,height=4.5cm}}
\caption{Elastic quark probing by polarized lepton.}
\label{sub1}
\end{minipage}
\hfill
\begin{minipage}{7.5cm}
\fbox{\epsfig{file=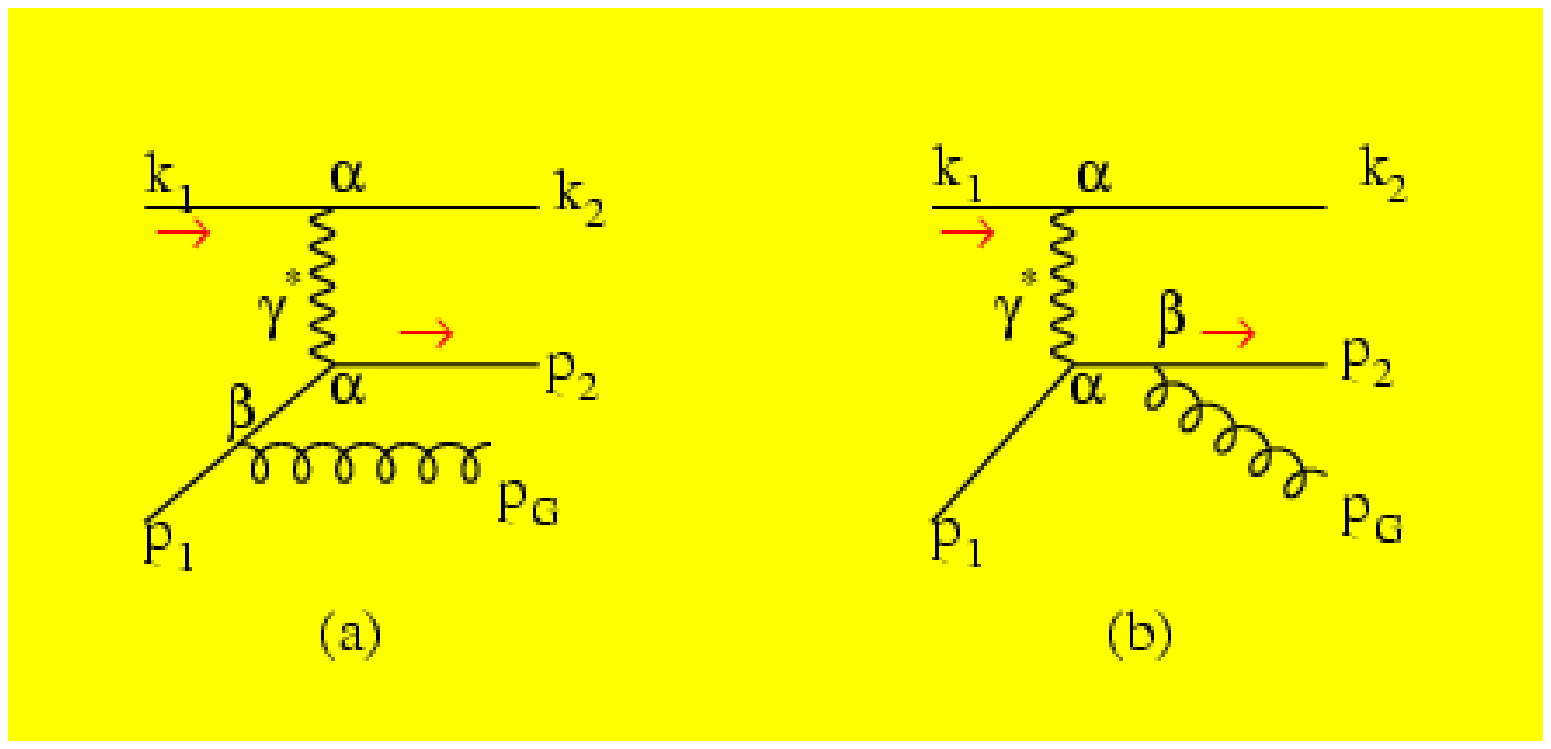,width=7.5cm,height=4.5cm}}
\caption{The gluon emission by the quark line.}
\label{sub2}
\end{minipage}
\end{figure}

The elementary subprocesses under study are the elastic quark
probing (Fig. \ref{sub1})
\begin{eqnarray}
  e^+_{\uparrow} ~+~ q \to e^+ ~+~ q_{\uparrow}' ~~~~({\rm LO})
\label{LO}
\end{eqnarray}
and the diagrams with  the gluon correction (Fig. \ref{sub2})
\begin{eqnarray}
  e^+_{\uparrow} ~+~ q \to e^+ ~+~ q_{\uparrow}' ~+~G ~~~~({\rm NLO}).
\label{NLO}
\end{eqnarray}

Since we are interested mainly in the central quark production
region, here we \\
neglect the contribution of the process
$e^+_{\uparrow} ~+~ G \to e^+ ~+~ q_{\uparrow} ~+~\bar q  $
with the quark pair production by $\gamma^{\ast}G$ coupling.
The spin transfer properties in the production of heavy $Q\bar Q$
quarks are discussed in ref.\,\cite{YuA}.

The purpose of this study is to look for kinematic regions
 where the polarization transfer \dLL gets enhanced in the 
subprocesses of fig.~1 and fig.~2. Being translated onto the
hadronic level, these kinematic regions, if any, can serve
as guidelines for the experiment HERMES.
The interval of DIS variables ($Q^2 ,\,\, y$ and other)
was chosen to be typical for HERMES.

The calculations are based on the following expression
for the scattered quark vector polarization 
$$
P_{out}^z = \frac{\left( 1+P_z^{in}\right)
\left( \vert M_{++} \vert^2 - \vert M_{-+} \vert^2 \right) +
\left( 1-P_z^{in}\right)
\left( \vert M_{+-} \vert^2 - \vert M_{--} \vert^2 \right) }
{\left( 1+P_z^{in}\right)
\left( \vert M_{++} \vert^2 + \vert M_{-+} \vert^2 \right) +
\left( 1-P_z^{in}\right)
\left( \vert M_{+-} \vert^2 + \vert M_{--} \vert^2 \right) }\, ,
$$
which can be derived from the final-state spin-density matrix.

\begin{figure}[h]
\begin{minipage}{6.5cm}
\fbox{\epsfig{file=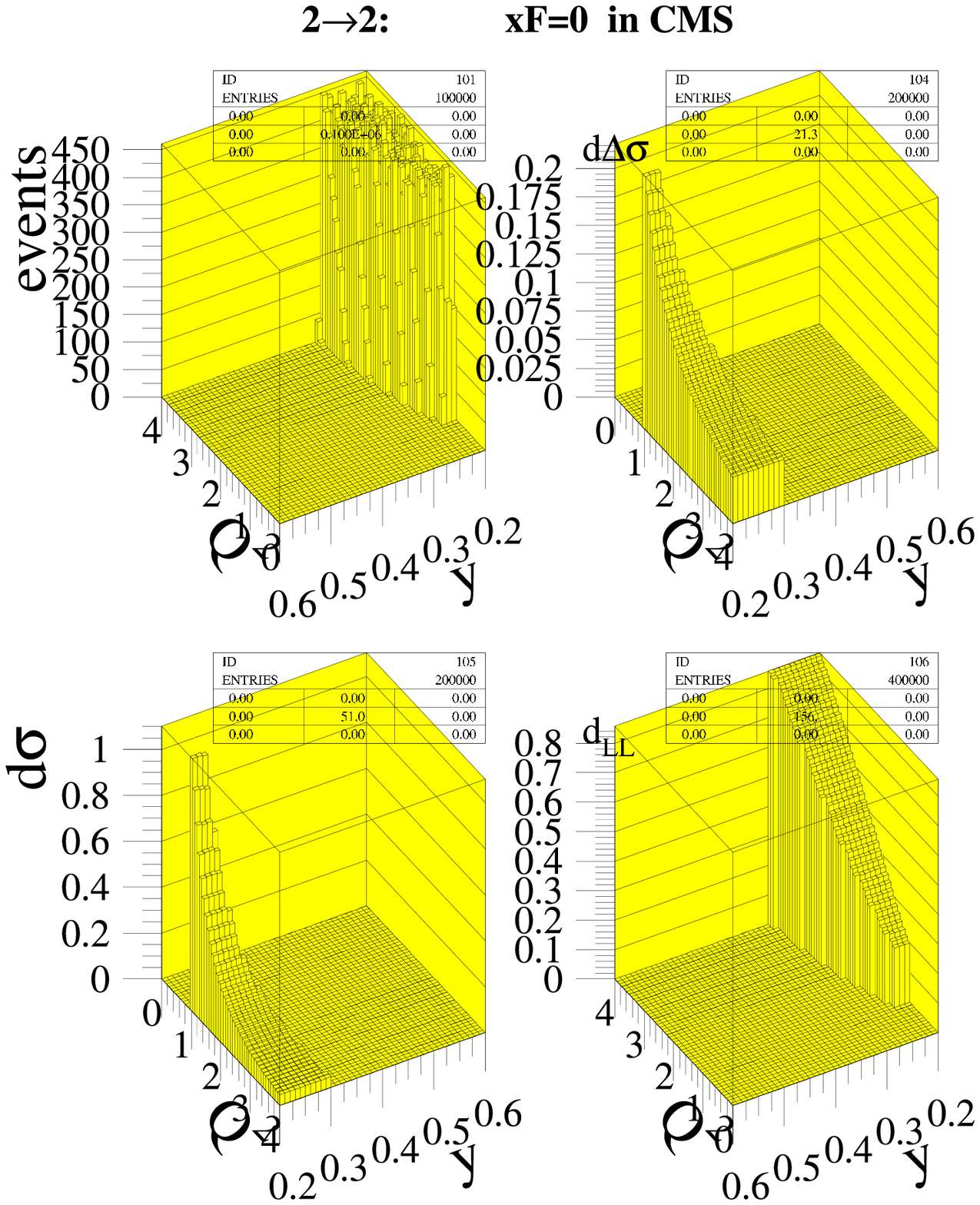,width=6.4cm,height=8.0cm}}
\caption{Elastic $\posup q\to e^+ \qup$ scattering.
($Q^2 ,y$) dependences of the
generated populations and differential cross section 
d$^2 \sigma$ (left column);  d$^2 \Delta\sigma$ 
and parameter $\hat d_{LL}$ (right column).}
\label{eqeq}
\end{minipage}
\hfill
\begin{minipage}{8.1cm}
\epsfig{file=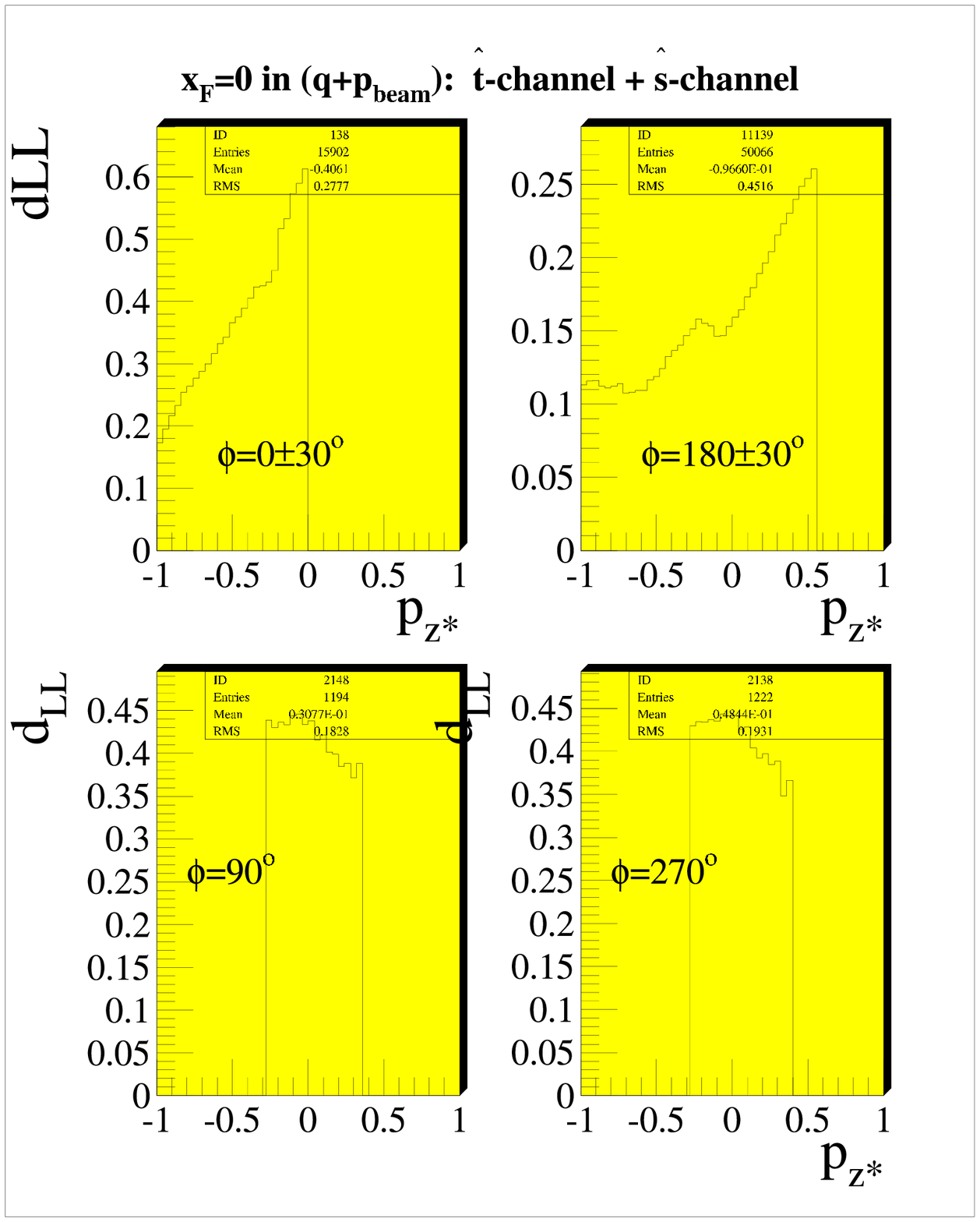,width=8.0cm,height=8.3cm}
\caption{The polarization transfer in the process 
$\posup q\to e^+ G \qup$ in the rest system of the
virtual photon and incoming quark. $p_z^{\ast}$ 
is the scattered quark momentum (in GeV). $\phi$ is
the angle between ($e^+ ,e^+$') and ($\gamma^{\ast},q'$) planes.
}
\label{eqeqg}
\end{minipage}
\end{figure}

The properties of the elastic subprocess (\ref{LO}) are plotted
in fig.~\ref{eqeq}. Both the differential cross section ${\rm d}\sigma$
and polarized differential cross section ${\rm d}\Delta\sigma$
decrease with increasing $Q^2$ so that the polarization transfer
\dLL increases rapidly. It achieves the value $\sim$0.8 at 
the extreme value of  $Q^2 \sim 4$\,GeV$^2$. For modest 
 momentum transfers, say $Q^2 \sim 2$\,GeV$^2$, \dLL has
the 'normal' values of $\sim$\,0.2.

Fig.\,\ref{eqeqg} represents the parameter \dLL calculated
for the reaction (\ref{NLO}) at four different values of 
the angle $\phi$ between two characteristic planes of the
reaction -- leptonic plane and the plane formed by the
virtual photon and the outgoing quark. As is known, the
polarization properties of the reactions 2\,$\to$\,3 depend
essentially, in general, on the rotational angle $\phi$.
The distributions of \dLL in fig.\,\ref{eqeqg} are shown for
four rotations of the planes with an increment equal to 90$^o$.
The lowest values of \dLL are obtained at $\phi$\,=\,180$^o$
when the normals to the said plains are anti-collinear.
The more sizeable values of \dLL appear when the normals make
right angles (90$^o$ or 270$^o$) -- up to $\sim$0.4.
And the most favorite kinematic region is predicted for 
the collinar normals ($\phi$\,=\,0) where the polarization 
transfer increases up to 0.6.

So coming down to the experiment (\ref{sidis}), one can expect
the sizeable longitudinal component of the $\Lambda$
 polarization induced by the
polarized positron under the kinematic configurations 
in which the normals to the leptonic plane and the plane
($\gamma^{\ast}$,\,$\Lambda$) are close to be collinear.
The numeric estimates made at the HERMES energy can certainly
be affected by the higher-order corrections.

\vfill


\end{document}